%% file: maintext.tex
\documentclass[%
  a4paper,
  amsfonts,amssymb,amsmath,
  reprint,
  superscriptaddress,        % <-- add this
  showkeys,nofootinbib,twoside
]{revtex4-2}
\usepackage[english]{babel}
\usepackage[utf8]{inputenc}
\usepackage{amsthm}
\usepackage{siunitx}
\usepackage[version=4]{mhchem}
\usepackage{comment}
\usepackage{graphicx}
\usepackage{subcaption}
\usepackage[colorinlistoftodos, color=green!40, prependcaption]{todonotes}
\usepackage{lineno}
\input{preamble}
\usepackage[pdftex, pdftitle={Article}, pdfauthor={Author}]{hyperref} % For hyperlinks in the PDF
\usepackage{ragged2e}
\begin{document}
\title{L-band single-photon generation from a cavity-coupled silicon C-center}

%\begin{comment}
\author{Carolina Crosta}
%\thanks{These authors contributed equally to this work.}
%\email{c.crosta3@campus.unimib.it}
\affiliation{Dipartimento di Scienza dei Materiali, Universit\`a di Milano-Bicocca and BiQute, via R. Cozzi, 20125 Milano, Italy}

\author{Kyu-Young Kim}
\affiliation{Institute for Research in Electronics and Applied Physics and Joint Quantum Institute, University of Maryland, College Park, Maryland 20742, USA}
\affiliation{Department of Electrical and Computer Engineering, University of Maryland, College Park, MD 20742, USA}
\affiliation{Department of Physics, Ulsan National Institute of Science and Technology, Ulsan 44919, Republic of Korea}

\author{Purbita Purkayastha}
%\email{herfati@umd.edu}
\affiliation{Institute for Research in Electronics and Applied Physics and Joint Quantum Institute, University of Maryland, College Park, Maryland 20742, USA}
\affiliation{Department of Physics, University of Maryland, College Park, Maryland 20742, USA}

\author{Abhijit Biswas}
%\email{abiswas3@umd.edu}
\affiliation{Institute for Research in Electronics and Applied Physics and Joint Quantum Institute, University of Maryland, College Park, Maryland 20742, USA}
\affiliation{Department of Electrical and Computer Engineering, University of Maryland, College Park, MD 20742, USA}

\author{Jasvith Raj Basani}
%\email{jasvith@umd.edu}
\affiliation{Institute for Research in Electronics and Applied Physics and Joint Quantum Institute, University of Maryland, College Park, Maryland 20742, USA}
\affiliation{Department of Electrical and Computer Engineering, University of Maryland, College Park, MD 20742, USA}

\author{Amirehsan Alizadehherfati}
%\email{herfati@umd.edu}
\affiliation{Institute for Research in Electronics and Applied Physics and Joint Quantum Institute, University of Maryland, College Park, Maryland 20742, USA}
\affiliation{Department of Electrical and Computer Engineering, University of Maryland, College Park, MD 20742, USA}

\author{Amirehsan Boreiri}
%\email{herfati@umd.edu}
\affiliation{Institute for Research in Electronics and Applied Physics and Joint Quantum Institute, University of Maryland, College Park, Maryland 20742, USA}
\affiliation{Department of Electrical and Computer Engineering, University of Maryland, College Park, MD 20742, USA}

\author{Chang-Min Lee}
\affiliation{Institute for Research in Electronics and Applied Physics and Joint Quantum Institute, University of Maryland, College Park, Maryland 20742, USA}
\affiliation{Department of Electrical and Computer Engineering, University of Maryland, College Park, MD 20742, USA}

\author{Fabio Pezzoli}
\email{fabio.pezzoli@unimib.it}
\affiliation{Dipartimento di Scienza dei Materiali, Universit\`a di Milano-Bicocca, BiQute and INFN-LNL, Via R. Cozzi, 55, Milano, 20125, Italy}

\author{Edo Waks}
\email{edowaks@umd.edu}
\affiliation{Institute for Research in Electronics and Applied Physics and Joint Quantum Institute, University of Maryland, College Park, Maryland 20742, USA}
\affiliation{Department of Electrical and Computer Engineering, University of Maryland, College Park, MD 20742, USA}

%\end{comment}
%\date{\today} % Leave empty to omit a date

\begin{abstract}
      Monolithic integration of single-photon sources into silicon photonics offers a compelling solution for realizing scalable quantum telecommunication architectures. Among the candidate silicon color centers, the C-center features a zero-phonon emission in the telecom L-band, corresponding to the spectral window of minimum transmission loss in optical fibers. Despite this advantage, isolating single C-centers has remained an unrealized challenge in solid-state quantum optics. In this work, we report the isolation of a single silicon C-center. We enhance the spontaneous emission rate of the zero-phonon line by coupling the single emitter to a nanophotonic crystal cavity, observing a fivefold reduction in the excited-state lifetime compared to bulk conditions, which corresponds to a Purcell factor lower bound of 33. Second-order autocorrelation measurements confirm single-photon emission from the investigated defect, obtaining an optical purity of $g^{(2)}(0) = 0.23$ without background correction. Our results establish a native telecom L-band quantum light source in silicon, laying the foundation for ultra-low-loss and long-distance quantum applications.
\end{abstract}

\maketitle

% INTRODUCTION

Color centers in silicon are rapidly emerging as the leading platform for integrated quantum photonic devices \cite{quard2025integration, sandholzer2026single, hollenbach2022wafer}. Unlike point defects in wide-bandgap materials such as diamond or silicon carbide, silicon-based emitters offer monolithic integration with mature, foundry-compatible silicon-on-insulator (SOI) photonics \cite{islam2026spin} while directly enabling emission in the optical telecommunication bands. The most studied carbon-based color centers in silicon are the W-, G-, T-, and C-centers \cite{davies2006radiation, andrini2024activation, quard2024femtosecond, khoury2022bright, chartrand2018highly, jhuria2024programmable, beaufils2018optical}. While the former three have been widely investigated at the single-defect level \cite{baron2022detection, redjem2020single, durand2024genuine, higginbottom2022optical, deabreu2022waveguide}, studies on C-centers have so far been limited to ensembles \cite{davies2006radiation, chartrand2018highly, wen2025optical, wagner1984excitation, wagner1984excitation}. In contrast to G- and T-centers that emit in the telecom O-band, the C-center offers a spectral advantage for long-range entanglement thanks to its native zero-phonon line in the telecom L-band (1565 - 1625 nm) \cite{khoury2022bright, zhu2026point}. This wavelength range coincides with the absolute transmission maximum of silica optical fibers and photonic waveguides \cite{udvarhelyi2022band}, making the C-center an ideal physical node for distributed photonic quantum computing and memories. 

\begin{figure}
    \centering
    \includegraphics[width=\columnwidth]{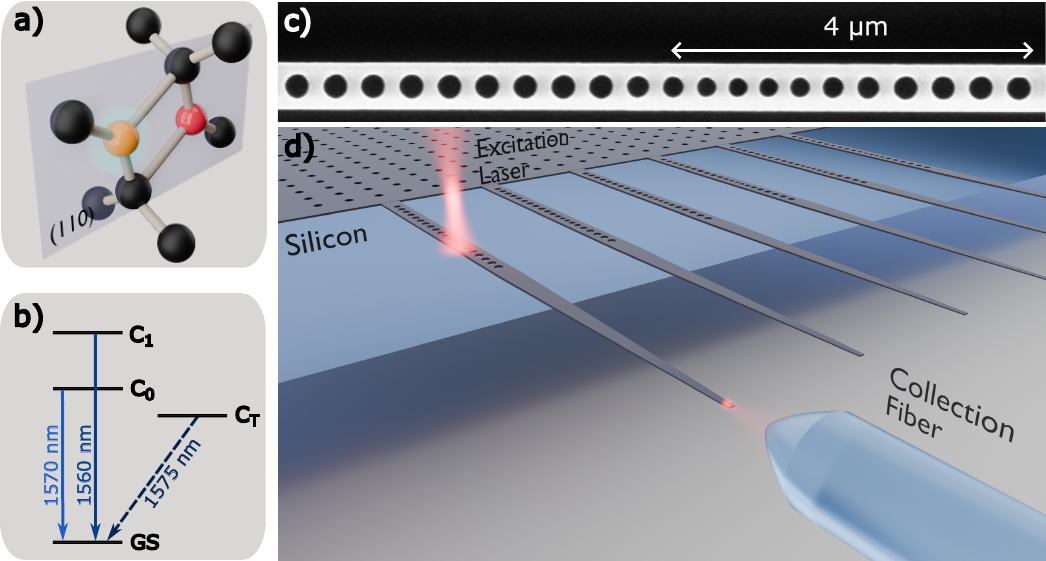}
    \caption{\textbf{a)} Atomic structure of the C-center. Black, orange and red spheres represent silicon, oxygen and carbon atoms respectively; \textbf{b)} Electronic structure of the C-center with transition wavelengths at cryogenic temperatures; \textbf{c)} SEM image of the nanophotonic crystal cavity; \textbf{d)} Schematic of the experimental configuration. The emitter is excited from top, while the emitted light is collected in-plane through a lensed single-mode fiber.}
    \label{fig:figure1}
\end{figure}

Despite their attractive properties, C-centers remain however unexplored at the single-emitter level. This silicon point defect consists of an interstitial carbon-interstitial oxygen pair (C\textsubscript{i}-O\textsubscript{i}) \cite{thonke1984carbon, davies2006radiation, chartrand2018highly}, shown in \textbf{Figure \ref{fig:figure1}a}. Both carbon and oxygen are natural silicon contaminants, which is another important feature of this defect. However, the presence of a lower-lying triplet state \cite{jones1973temperature, wen2025optical} (see \textbf{Figure \ref{fig:figure1}b}) and recombination into the incoherent phonon sideband result in a slow and inefficient zero-phonon transition, thus hindering the observation of single emitters. A platform that enhances the brightness of C-centers while enabling the spatial and spectral isolation of individual emitters is thus strongly needed. High-quality-factor nanophotonic cavities provide both capabilities. Through the Purcell effect, they enhance the spontaneous emission rate within a narrow spectral window, while their tightly confined optical modes allow to selectively address individual emitters \cite{gritsch2023purcell}. This approach has already enabled the identification and spontaneous emission rate enhancement of individual W- \cite{lefaucher2023purcell}, G- \cite{lefaucher2023cavity, kim2025bright}, and T-centers \cite{johnston2024cavity, islam2023cavity}, but has not yet been extended to C-centers.

In this work, we report the isolation of a single silicon C-center via coupling to a photonic crystal cavity in a tapered nanobeam waveguide. The small mode volume of the photonic crystal cavity enables accessing C-centers at a single-defect level. Furthermore, it selectively increases the spontaneous emission rate, resulting in a significant enhancement of the zero-phonon line emission. We show a lifetime reduction of five times, reaching \SI{0.287}{\micro\second}, which corresponds to a Purcell factor $F_\mathrm{P} \ge 33$. This cavity-enhanced emission enables the direct observation of single-photon generation from a C-center, extracting a zero-delay second-order correlation $g^{(2)}(0) = 0.23$ without background correction. These results establish a route toward bright and individually addressable telecom L-band emitters in silicon, opening new possibilities to employ C-centers in scalable quantum computing and communication architectures.

\begin{figure}
    \centering
    \includegraphics[width=\columnwidth]{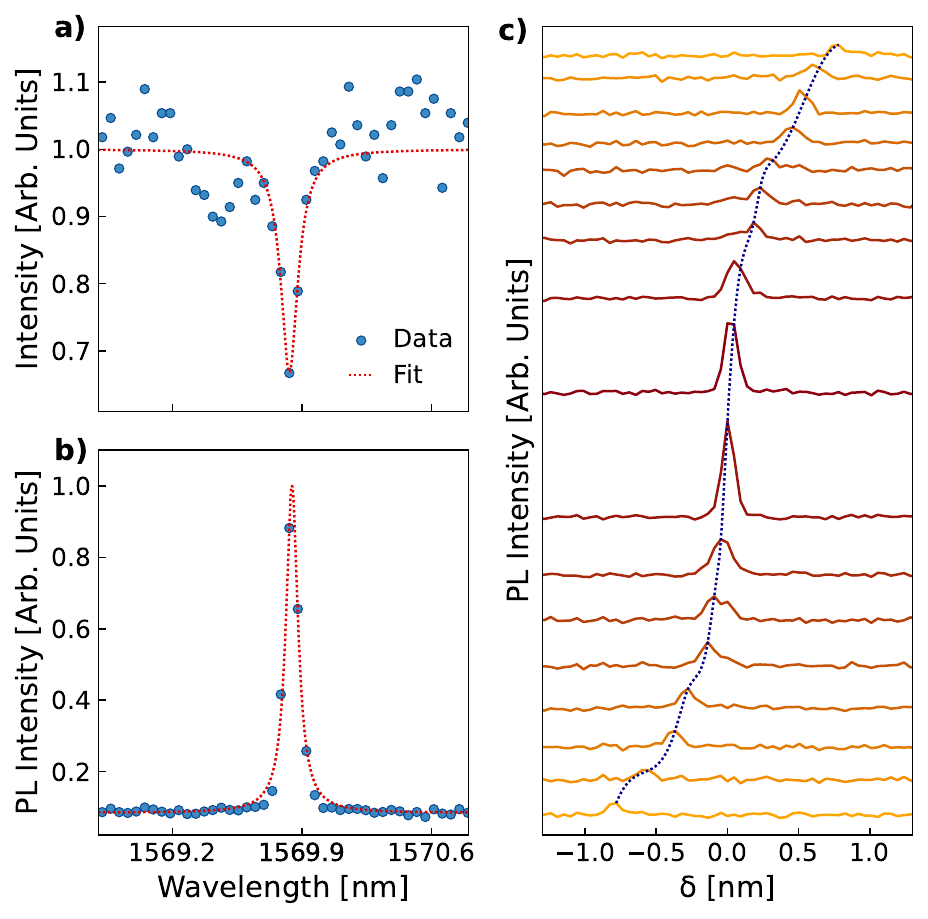}
    \caption{\textbf{a)} Reflectivity spectrum of the cavity tuned at the C\textsubscript{0}-line wavelength (dots) and corresponding Lorentzian fit (dashed line), extracting $\text{FWHM} = 0.11 \pm 0.02$ nm that corresponds to a quality factor of $(1.4 \pm 0.3) \times 10^4$; \textbf{b)} Photoluminescence spectrum (dots) of the emitter at \SI{8}{\kelvin} under \SI{800}{\nano\meter} CW excitation, pump power of \SI{10}{\micro\watt}, in on-resonance condition. The Lorentzian fit of the experimental data (dashed line) provides $\lambda_{\mathrm{ZPL}} = 1569.831 \pm 0.004$ nm and $\text{FWHM} = 0.078 \pm 0.001$ nm; \textbf{c)} Photoluminescence intensity as a function of the cavity mode detuning $\delta$. The dashed line is a visual guide to follow the change of $\delta$.}
    \label{fig:figure2}
\end{figure}

% CENTER STRUCTURE + EXPERIMENTAL DETAILS
The C\textsubscript{i}--O\textsubscript{i} complex in silicon features a zero-phonon line at \SI{1570}{\nano\meter} (C\textsubscript{0}-line), characterized by a lifetime in the \SI{}{\micro\second} regime \cite{thonke19850, bohnert1993transient}. The second excited state emits at \SI{1560}{\nano\meter} (C\textsubscript{1}-line) \cite{wagner1984excitation}, while a long-living dark triplet state (C\textsubscript{T}) is located \SI{2.64}{\milli\electronvolt} below the C\textsubscript{0}-state \cite{ishikawa2009photoluminescence, ishikawa2011optical} (see \textbf{Figure \ref{fig:figure2}b}). The C\textsubscript{T}-state lifetime at cryogenic temperatures is in the order of \SI{}{\milli\second} and the C\textsubscript{0}-C\textsubscript{T} intersystem crossing lifetime is expected to be $\tau_\mathrm{ISC}\sim$\SI{2.8}{\micro\second} \cite{bohnert1993transient}, contributing to the C-center poor emission efficiency. The theoretically predicted Debye-Waller factor for the C-center ranges between 0.12 \cite{udvarhelyi2022band} and 0.2 \cite{silkinis2025optical}, while experimental estimations are not reported in literature to date. The combination of such low Debye-Waller factor and the presence of dark long-living states, makes it necessary to selectively enhance the C-center zero-phonon transition.

In this work C-centers are generated in \textsuperscript{28}Si isotopically and chemically purified SOI to minimize impurity concentration, following the recipe previously employed to create T-centers \cite{lee2023high}. Further details on the implantation conditions and the generation of C-centers are reported in \textbf{Section \ref{Supp:implantation}} and \textbf{\ref{Supp:C1-line}} of the Supplementary Material. Carbon and hydrogen implantation were performed with a tilting angle of \SI{7}{\degree} to avoid channeling phenomena, with an energy corresponding to an ion projected range of \SI{110}{\nano\meter} from Stopping and Range of Ions in Matter (SRIM) simulations. We exploit natural oxygen contamination of silicon and the possible diffusion of oxygen atoms from the silicon oxide layer during implantation and annealing processes as source of oxygen to generate the C\textsubscript{i}-O\textsubscript{i} complex of interest.

The implanted SOI is fabricated into tapered nanobeams containing a one-dimensional photonic crystal cavity \cite{islam2023cavity, biswas2025single} (see \textbf{Figure \ref{fig:figure1}c}), which is resonant with the C\textsubscript{0} transition. The cavity mode is finely tuned at the C-center emission wavelength by condensing nitrogen gas on the device surface \cite{kim2025bright}. The detailed explanation of the cavity design can be found in \textbf{Section \ref{Supp:cavity}} of the Supplementary Material. The number of holes at the left ($N_L$) and right ($N_R$) edges of the cavity can be independently selected to obtain asymmetric reflectance of the photonic crystal cavity, allowing directional photon propagation along the waveguide towards the tapered regime, thus maximizing the coupling of the guided mode to the lensed single-mode fiber. The experimental configuration is depicted in \textbf{Figure \ref{fig:figure1}d}. 

The sample is characterized in a closed-cycle cryostat, reaching a base temperature of \SI{8}{\kelvin}. The C-center is excited along the direction perpendicular to the sample surface by a continuous-wave (CW) \SI{800}{\nano\meter} or a pulsed \SI{780}{\nano\meter} laser. The pump laser is focused on the sample using a 100x objective lens with a NA of 0.70. The signal is collected through a lensed single-mode fiber with a NA of 0.4, reaching $\sim$70\% coupling efficiency, and detected with a spectrometer or in-fiber spectral filtering coupled to a superconducting nanowire single-photon detector. \textbf{Section \ref{Supp:exp_setup}} of the Supplementary Material provides further details on the experimental setup.

% RESULTS 1: reflectivity and PL vs detuning

To characterize the nanophotonic crystal cavity, we tune the cavity mode at the expected C\textsubscript{0}-line wavelength and perform direct reflectivity measurements. By shining through the lensed fiber the light of a tungsten-halogen lamp, we measure the reflected spectrum of the cavity mode, shown in \textbf{Figure \ref{fig:figure2}a}. From Lorentzian fitting of the experimental data, we extract a reflected cavity mode full width at half maximum (FWHM) of $0.11 \pm 0.02$ nm, corresponding to a quality factor of $(1.4 \pm 0.3) \times 10^4$. 

Next, we excite the center of the cavity with a \SI{10}{\micro\watt} \SI{800}{\nano\meter} CW laser and measure the C\textsubscript{0}-line photoluminescence spectrum in on-resonance condition. The experimental data (\textbf{Figure \ref{fig:figure2}b}) exhibit a single photoluminescence peak, corresponding to the C-center zero-phonon line. Through Lorentzian fit of the spectrum (dashed line), we extract the peak center $\lambda_{\mathrm{ZPL}} = 1569.831 \pm 0.004$ nm, and linewidth $\mathrm{FWHM} = 0.078 \pm 0.001$ nm, which is slightly narrower than that of the cavity mode.

To evaluate the cavity-emitter interaction, we track the zero-phonon line intensity as a function of the cavity detuning $\delta = \lambda_{\mathrm{ZPL}}-\lambda_{\mathrm{CM}}$, where $\lambda_{\mathrm{CM}}$ is the cavity mode wavelength obtained from independent reflected spectrum measurements. \textbf{Figure \ref{fig:figure2}c} displays the photoluminescence spectra over a range of cavity detunings $\delta$, showing a fast decrease of the C\textsubscript{0}-line intensity as the cavity mode is spectrally shifted from $\lambda_{\mathrm{ZPL}}$. Through Lorentzian fit of the C\textsubscript{0}-line integrated area as a function of $\delta$, we extract $\mathrm{FWHM} = 0.13 \pm 0.01$ nm. This result is comparable to the value obtained from the reflectivity spectrum, thus confirming that the photoluminescence enhancement of the C-center arises from the cavity coupling. Details on this analysis can be found in \textbf{Section \ref{Supp:PLvsDET}} of the Supplementary Material.

% RESULTS 2: lifetime measurements vs detuning

\begin{figure}
    \centering
    \includegraphics[width=\columnwidth]{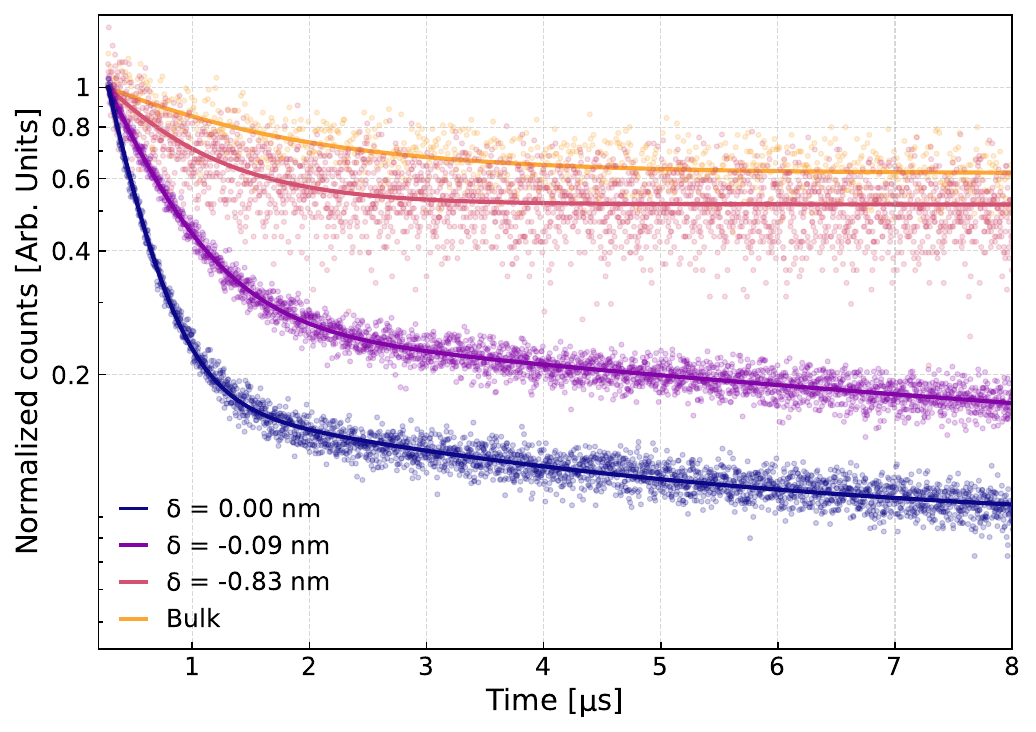}
    \caption{Time-resolved photoluminescence measurements of a C-center ensemble in a bare nanobeam (orange curve), providing $\tau_0 = 1.45 \pm 0.08$ \SI{}{\micro\second} and a single C-center in a cavity tuned (i) on-resonance ($\delta = 0$ \SI{}{\nano\meter},  $\tau_{\mathrm{on}} = 0.287 \pm 0.002$ \SI{}{\micro\second}); (ii) slightly off-resonance ($\delta = -0.09$ \SI{}{\nano\meter}, showing a fast increase of the lifetime to $\tau = 0.523 \pm 0.005$ \SI{}{\micro\second}); and (iii) off-resonance ($\delta = -0.83$ \SI{}{\nano\meter}, $\tau_{\mathrm{off}} = 0.780 \pm 0.030$ \SI{}{\micro\second}). The C-center is measured at \SI{8}{\kelvin} and excited by a \SI{780}{\nano\meter} pulsed laser with a repetition rate of \SI{100}{\kilo\hertz}.}
    \label{fig:figure3}
\end{figure}

We then perform time-resolved photoluminescence measurements (\textbf{Figure \ref{fig:figure3}}) to quantify the Purcell enhancement of the C-center. The emitter is excited by a \SI{780}{\nano\meter} pulsed laser with a repetition rate of \SI{100}{\kilo\hertz} and a pulse width of \SI{70} {\pico\second}, while keeping the fiber filter central wavelength fixed at $\lambda_\mathrm{ZPL}$. As a reference, we measure the ensemble lifetime $\tau_0 = 1.45 \pm 0.08$ \SI{}{\micro\second} in a nanobeam without cavity, i.e., with only a Bragg mirror to redirect the emitted light towards the lensed fiber. This value is in good agreement with previous reports in bulk silicon \cite{thonke19850}. Then, we characterize the lifetime of a C-center in a cavity tuned on-resonance ($\delta = 0$ \SI{}{\nano\meter}), deriving a lifetime $\tau_{\mathrm{on}} = 0.287 \pm 0.002$ \SI{}{\micro\second}. This time-resolved photoluminescence result determines a fivefold reduction of the C-center lifetime compared to that of the ensemble.

Under slightly off-resonance ($\delta = -0.09$ \SI{}{\nano\meter}) condition, the lifetime shows a fast increase to $\tau = 0.523 \pm 0.005$ \SI{}{\micro\second}, while off-resonance detuning ($\delta = -0.83$ \SI{}{\nano\meter}) is characterized by $\tau_{\mathrm{off}} = 0.780 \pm 0.030$ \SI{}{\micro\second}. The $\tau$ values obtained at different $\delta$ are provided in \textbf{Section \ref{Supp:TAUvsDET}} of the Supplementary Material, confirming the cavity-driven nature of the spontaneous emission rate enhancement. The difference between $\tau_{\mathrm{off}}$ and $\tau_0$ has been reported both for G- and T-centers coupled to a nanophotonic crystal cavity, and can be attributed to additional non-radiative recombination pathways present in the nanobeam \cite{kim2025bright, johnston2024cavity}. We also observe that for small detuning conditions, the experimental data are well described by a double-exponential fit. We attribute the fast decay component to the C-center coupled to the cavity, and the long-lifetime component, in the order of few \SI{}{\micro\second}, to background emission originating from emitters uncoupled to the cavity.

From the above lifetime measurements, we can estimate $F_\mathrm{P}$ associated with the emitter-cavity coupling via the equation \cite{kim2025bright}:
\begin{equation}
    F_\mathrm{P} = \tau_0 \left( \frac{1}{\tau_{\mathrm{on}}} - \frac{1}{\tau_{\mathrm{off}}} \right) \cdot \frac{1}{F_{\mathrm{DW}}} \cdot \frac{1}{\eta_{\mathrm{QE}}}
\end{equation}
where $F_{\mathrm{DW}}$ is the C-center Debye-Waller factor, and $\eta_{QE}$ is the emitter internal quantum efficiency. If we consider the intersystem crossing to be the intrinsic non-radiative decay process of the C-center, then $\eta_{QE} = \frac{\Gamma_{0} - \sum \Gamma_{\mathrm{nr}}}{\Gamma_{0}} \leq \frac{\Gamma_{0} - \Gamma_{\mathrm{ISC}}}{\Gamma_{0}} = 0.482$; where $\Gamma_{0} = 1/\tau_{0}$ is the total decay rate including radiative and non-radiative decay processes, $\Gamma_{nr}$ is the decay rate of each non-radiative decay process, and $\Gamma_{\mathrm{ISC}}$ is the intersystem crossing rate. By substituting the previously computed lifetime values, $\eta_{\mathrm{QE}} = 0.482$, and $F_{\mathrm{DW}} = 0.2$ as the highest Debye-Waller factor bound from theoretical predictions, we estimate $F_\mathrm{P} \ge 33$. Additional non-radiative decay channels different from the intersystem crossing process would further reduce $\eta_{\mathrm{QE}}$, leading to even larger $F_{\mathrm{P}}$ values.

% RESULTS 3: saturation curve and g2

\begin{figure}
    \centering
    \includegraphics[width=\columnwidth]{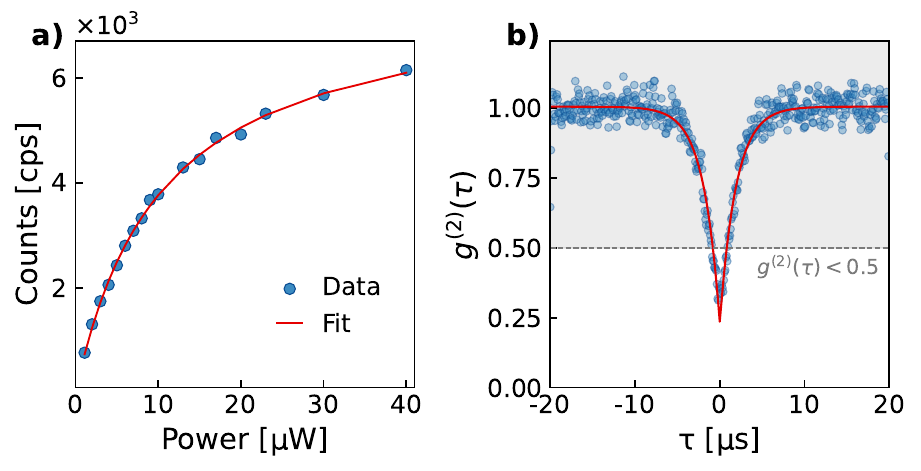}
    \caption{\textbf{a)} On-resonance C-center count rate as a function of \SI{800}{\nano\meter} CW pump power (dots) and two-level system fit (solid line) with \textbf{Equation \ref{eq:saturation}}, extracting $P_{\mathrm{sat}} = 10.4 \pm 0.3$ \SI{}{\micro\watt} and $I_{\mathrm{sat}} = (7.68  \pm 0.09) \times 10^3$ cps; \textbf{b)} Second-order autocorrelation measurement of a C-center under \SI{7}{\micro\watt} = $0.67P_{\mathrm{sat}}$ at \SI{800}{\nano\meter} CW excitation (dots) and corresponding fit (solid line) with \textbf{Equation \ref{eq.g2}}, extracting $g^{(2)}(0) = 0.23 \pm 0.01$.}
    \label{fig:figure4}
\end{figure}

Having engineered a highly efficient emission regime, we next evaluate the system viability as a deterministic quantum light source starting from the investigation of the emitter count rate as a function of the pump power. The experimental data are shown as dots in \textbf{Figure \ref{fig:figure4}a}, while the solid line represents a fit via a two-level system model, which is described by the equation \cite{aharonovich2010photophysics}:
\begin{equation}\label{eq:saturation}
    I(P) = \frac{I_{\mathrm{sat}} \cdot P}{P + P_{\mathrm{sat}}}
\end{equation}
where $I_{\mathrm{sat}}$ is the saturation intensity and $P_{\mathrm{sat}}$ is the saturation power of the emitter. From the fit we extract $I_{\mathrm{sat}} = (7.68  \pm 0.09) \times 10^3$ cps and $P_{\mathrm{sat}} = 10.4 \pm 0.3$ \SI{}{\micro\watt}. This behavior confirms that this emitter is well described by a single two-level model.

To finally prove the single-emitter nature of the studied defect, we perform second-order autocorrelation measurements below the saturation power level. \textbf{Figure \ref{fig:figure4}b} shows $g^{(2)}(\tau)$ of the center (dots) under \SI{800}{\nano\meter} CW excitation with pump power of \SI{7}{\micro\watt} = $0.67P_{\mathrm{sat}}$ and corresponding fit according to the following equation \cite{kim2013ultrafast}:  
\begin{equation} \label{eq.g2}
    g^2(\tau) = A \cdot \left[ 1 - (1-g^2(0)) \cdot \exp\left( - \frac{|\tau-t_0|}{\tau_C}\right) \right]
\end{equation}
where $g^{(2)}(0)$ is the anti-bunching value at zero-delay of the second-order correlation function, and $\tau_C$ is the cavity-coupled emitter lifetime. Hence, from the experimental data we extract $g^{(2)}(0) = 0.23 \pm 0.01$ without background correction. The $g^{(2)}(0)$ values obtained at different excitation powers are reported in \textbf{Figure \ref{fig:Supp_g2(0)}} of the Supplementary Material, further confirming that this emitter is a single C-center.

% CONCLUSIONS

In conclusion, we report the isolation of a single silicon C-center. To selectively enhance the spontaneous emission rate of the C-center zero-phonon line, we couple the emitter to a high quality factor nanophotonic crystal cavity in a tapered nanobeam waveguide. We observe a fivefold lifetime reduction of the C-center from \SI{1.45}{\micro\second} in bulk to \SI{0.287}{\micro\second} in on-resonance conditions of the cavity mode. The increased defect photoluminescence when in resonance with the cavity mode allows second-order autocorrelation measurements, eventually demonstrating single photon emission with $g^{(2)}(0) = 0.23$ without background correction. While future optimization of the C-center generation recipe and precisely scaling defect concentration will further improve the signal-to-noise ratio, our nanophotonic architecture already establishes the C-center as a highly viable quantum node. The generation of single photons in the telecom L-band within a foundry-compatible SOI platform, provides a scalable route to employing color centers in future long-range, distributed quantum networks.

% DECLARATIONS AND AUTHOR CONTRIBUTIONS

\section*{Acknowledgments}
The authors acknowledge Yuxi Jiang, Fariba Islam, and Jacopo Pedrini for valuable discussion. C.C. acknowledges the Ermenegildo Zegna Founder's Scholarship for financial support. The Waks' group acknowledges financial support from the National Science Foundation (grant \#ECCS2423788), the Department of Energy (grant \#DESC0026071), and the Air Force Office of Scientific Research (grant \#FA95502310667 and \#FA95502410266).

\section*{Conflict of interests}
The authors declare no competing interests.

\section*{Data availability}
The data of this work are included in the published article. Additional raw data are available from the corresponding authors upon request. 

\section*{Author contributions}
F.P. devised the project. C.C., P.P., F.P., and E.W. conceived the experiment. K.K., Ab.B., and Am.B. helped setting up the experiment. P.P. fabricated the device. C.C. performed the experiments and analyzed the data. J.R.B. and A.A. assisted with preparation of the figures and data analysis. C.C., K.K., C.L., F.P., and E.W. prepared the manuscript. K.K., C.L. and E.W. supervised the experimental activities. All authors discussed the results and confirmed the manuscript.

\newpage
\bibliographystyle{ieeetr}   % or apalike, plain, unsrt, etc.
\bibliography{refs.bib}          % assumes refs.bib is in the same folder
%\nocite{*}

\clearpage
\onecolumngrid
\section*{Supplementary Material}
\setcounter{subsection}{0}
\renewcommand{\thesubsection}{S\arabic{subsection}}
\setcounter{figure}{0}
\renewcommand{\thefigure}{S\arabic{figure}}
\setcounter{equation}{0}
\renewcommand{\theequation}{S\arabic{equation}}

\subsection{Implantation conditions} \label{Supp:implantation}
C-centers are generated in SOI employing the following recipe: carbon implantation (energy of \SI{38}{\kilo\electronvolt}, dose of $7\times10^{12}$ ions/cm\textsuperscript{2}) is followed by a rapid thermal annealing step at \SI{1000}{\celsius} for \SI{20}{\second} in argon atmosphere, and subsequent hydrogen implantation (energy of \SI{9}{\kilo\electronvolt}, dose of $7\times10^{12}$ ions/cm\textsuperscript{2}) is followed by annealing at \SI{400}{\celsius} for \SI{3}{\minute} in nitrogen atmosphere.

\subsection{Photoluminescence of a bare nanobeam}\label{Supp:C1-line}

\begin{figure}[h!]
    \centering
    \includegraphics[width=0.5\linewidth]{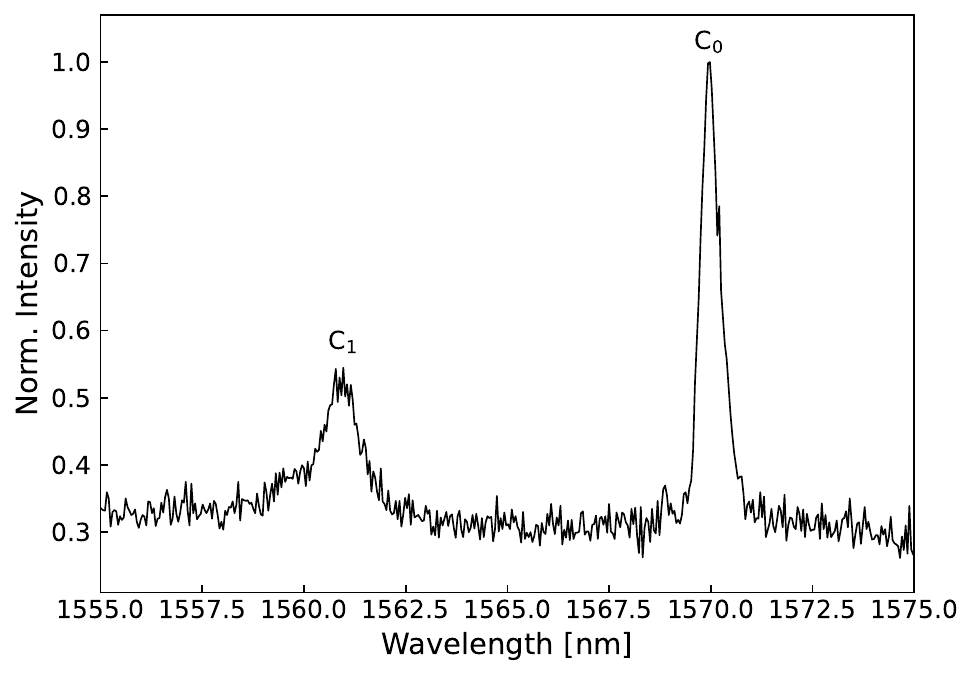}
    \caption{Photoluminescence spectrum of an ensemble of C-centers in a bare nanobeam, showing the C\textsubscript{0}-line at $\sim$\SI{1570}{\nano\meter} and the C\textsubscript{1}-line at $\sim$\SI{1560}{\nano\meter}.}
    \label{fig:supp_C0_C1}
\end{figure}

To prove that C-centers are generated in SOI by our implantation and annealing procedure, we perform photoluminescence measurements of a bare nanobeam, i.e. with only a Bragg mirror to redirect the emitted light toward the tapered end of the structure. \textbf{Figure \ref{fig:supp_C0_C1}} illustrates the resulting photoluminescence spectrum, showing the zero-phonon line at $\sim$\SI{1570}{\nano\meter} and the emission from the second excited state at $\sim$\SI{1560}{\nano\meter}. The presence of this pair of photoluminescence peaks is a fingerprint of the C-center, confirming that we are effectively generating ensembles of C-centers. The exact same SOI sample is then fabricated into nanobeams containing nanophotonic crystal cavities, thus implying that the single emitter studied in this paper is a C-center. 

\subsection{Cavity design} \label{Supp:cavity}

To enhance the radiative emission rate of C-centers, we couple them to a one-dimensional photonic crystal cavity within an adiabatic tapered silicon nanobeam waveguide. This tapered regime allows to couple the nanobeam to a lensed single-mode fiber with NA of 0.4. 

The photonic crystal cavity consists of a row of air holes with radius $r_i$ and distance $a_i$. The cavity is obtained by linearly tapering $r_i$ from \SI{144}{\nano\meter} to \SI{105}{\nano\meter} to tune its resonance at the C-center zero-phonon transition of $\sim$\SI{1570}{\nano\meter}. The simulated cavity mode volume is $0.1 \left( \frac{\lambda}{n} \right) ^3$, where $\lambda$ is the resonant wavelength of the cavity mode and n is the refractive index of the cavity material. The number of holes at each edge of the cavity can be independently controlled. To favor light emission from the waveguide taper into the lensed single-mode fiber, we employed $N_L = 13$ at the non-tapered end and $N_R = 7$ at the tapered end of the nanobeam. 

The nanobeam has width $b$, which decreases to $b_{\mathrm{taper}}$ over the adiabatic taper length $L_{\mathrm{taper}}$. From finite-difference time-domain simulations the optimized device parameters are set to $b=\SI{510}{\nano\meter}$, $b_{\mathrm{taper}}=\SI{110}{\nano\meter}$, and $L_{\mathrm{taper}}=\SI{13}{\micro \meter}$.

The nanobeams are fabricated from the implanted SOI wafer via electron-beam lithography and subsequent etching. After fabrication, the devices are transferred using a polydimethylsiloxane stamp to pick up individual pads with nanobeams and place them at the edge of a silicon carrier wafer, leaving the nanobeam in a suspended configuration.

\subsection{Experimental setup} \label{Supp:exp_setup}

The sample is loaded in a closed-loop cryostat (S50 Montana Instruments) reaching a base temperature of \SI{8}{\kelvin} equipped with a lensed fiber probe station. The emitter is excited from top using a 100x objective with NA of 0.70 (Mitutoyo, Plan Apo NIR). CW above-band excitation at \SI{800}{\nano\meter} (Msquared, Solstis) is employed for photoluminescence and second-order autocorrelation measurements, while time-resolved photoluminescence measurements are carried out exciting with a \SI{780}{\nano\meter} pulsed laser (Picoquant, LDH-D-C-780) externally triggered to a repetition rate of \SI{100}{\kilo\hertz}.

The nanobeam is coupled to a lensed single-mode fiber with a working distance of \SI{14}{\micro\meter} by measuring and maximizing the reflected light power of a CW superluminescent diode at \SI{1566}{\nano\meter} (Thorlabs), reaching coupling efficiencies of $\sim$70\%. For photoluminescence measurements the collected signal is detected by a Princeton Instruments spectrometer, comprising a monochromator (600\,g/mm and 150\,g/mm grating) and a 1024 pixel InGaAs array. For time-resolved photoluminescence and second-order autocorrelation measurements the zero-phonon line signal is filtered by a tunable fiber-coupled filter (WLPhotonics) with a bandwidth of \SI{0.2}{\nano\meter}. The filtered signal is then split by a \(50{:}50\) fiber-coupled beam splitter (Thorlabs) for correlation measurements and detected via fiber-cupled superconducting nanowire single-photon detector (QuantumOpus) using a high speed time tagger (HydraHarp 400).

\subsection{Photoluminescence as a function of cavity mode detuning} \label{Supp:PLvsDET}

\begin{figure}[h!]
    \centering
    \includegraphics[width=0.5\linewidth]{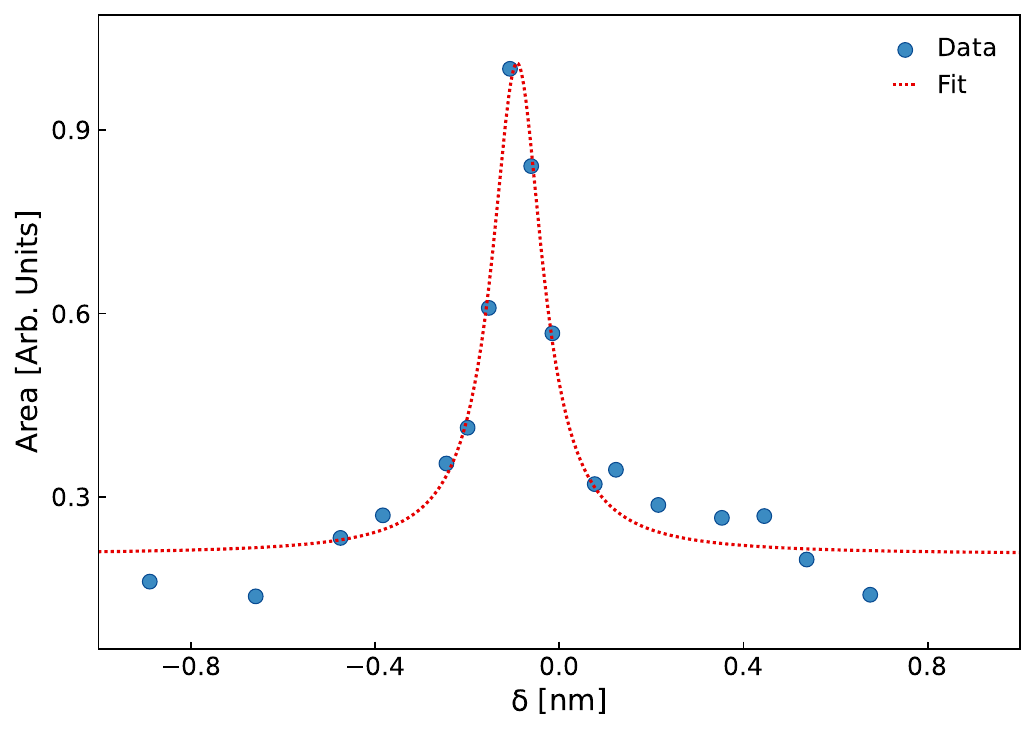}
    \caption{Integrated area of the C\textsubscript{0}-line as a function of the cavity mode detuning $\delta$ from \textbf{Figure \ref{fig:figure2}c} (dots). The dashed line represents the Lorentzian fit of the experimental data, extracting a FWHM of $0.13 \pm 0.01$ \SI{}{\nano\meter} and $\lambda_C = 1569.831 \pm 0.004$ \SI{}{\nano\meter}.}
    \label{fig:Supp_area}
\end{figure}

\textbf{Figure \ref{fig:Supp_area}} shows the C\textsubscript{0}-line integrated area as a function of the cavity mode detuning $\delta$ from \textbf{Figure \ref{fig:figure2}c} (dots). Via Lorentzian fit of the experimental data (dashed line), we extract a FWHM of $0.13 \pm 0.01$ nm. This value is comparable to the cavity mode FWHM obtained from reflectivity measurements, confirming that the photoluminescence enhancement of the C\textsubscript{0}-line is induced by cavity coupling. The fit also provides a central wavelength $\lambda_C = 1569.831 \pm 0.004$ nm, which we use as the on-resonance condition for lifetime and second-order autocorrelation measurements.

\subsection{Lifetime as a function of cavity mode detuning} \label{Supp:TAUvsDET}

\begin{figure}[h!]
    \centering
    \includegraphics[width=0.5\linewidth]{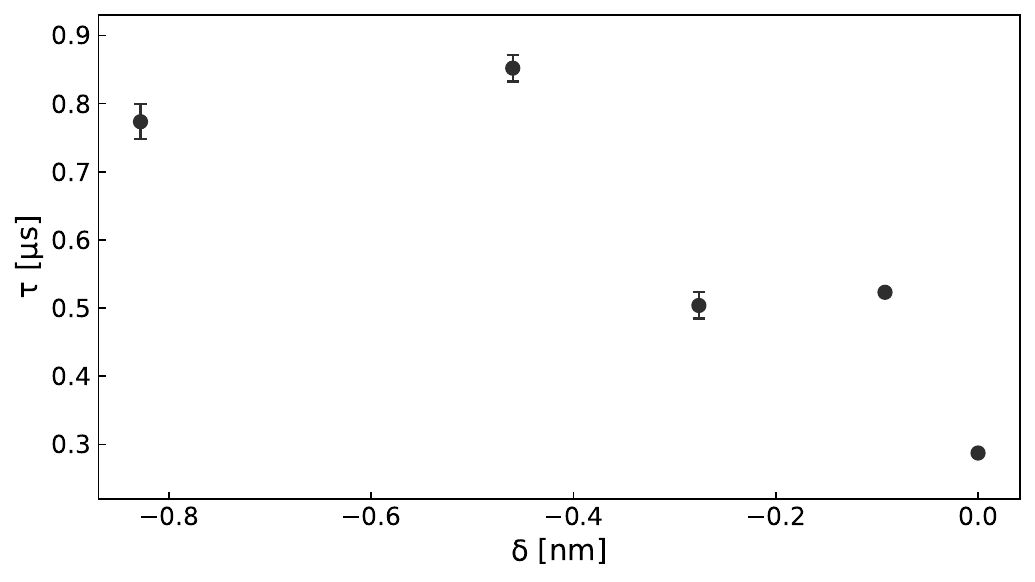}
    \caption{C-center lifetime as a function of the cavity mode detuning $\delta$.}
    \label{fig:Supp_lifetime}
\end{figure}

\textbf{Figure \ref{fig:Supp_lifetime}} shows the lifetime extracted from single- or double-exponential fit of the experimental data of the cavity-coupled C-center as a function of the detuning $\delta$. The lifetime increases with $\delta$, proving that the spontaneous emission rate enhancement is induced by the cavity.

The off-resonance lifetime $\tau_{\mathrm{off}} = 0.780 \pm 0.030$ is shorter than the lifetime of an ensemble of C-centers in a bare nanobeam $\tau_0 = 1.45 \pm 0.08$. This difference has been previously reported both for G- and T-centers integrated into a nanophotonic crystal cavity \cite{kim2025bright, johnston2024cavity}, and it can be ascribed to additional non-radiative decay paths caused by e.g. etched sidewalls of the cavity architecture. 

We finally estimate the Purcell factor to be $F_\mathrm{P} \ge 33$. From simulations of the quality factor and the cavity mode volume, the theoretical Purcell factor is $F_\mathrm{P,sim} = 850$. Experimentally, this value is lowered by position and orientation mismatch of the emitter with respect to the cavity field maximum and polarization.

\subsection{\texorpdfstring{$g^{(2)}(0)$}{g(2)(0)} as a function of pump power}

We perform second-order autocorrelation measurements at three different powers: (i) low power (\SI{4}{\micro\watt}), (ii) middle power (\SI{7}{\micro\watt}), and (iii) close to saturation power (\SI{10}{\micro\watt}). The experimental data are fitted with the following equation \cite{kim2013ultrafast}:
\begin{equation} \label{eq.g2_bg}
    g^{(2)}(\tau) = A \cdot \left( 1 - \left( \frac{I_S-I_B}{I_S} \right)^2 \cdot (1-g^{(2)}(0)) \cdot \exp\left( - \frac{\tau-t_0}{\tau_C}\right) \right)
\end{equation}
where A is a pre-factor accounting for imperfect normalization of the experimental data, $I_S$ is the signal count rate, $I_B$ is the background count rate acquired by keeping the fiber filter at $\lambda_{ZPL}$ and detuning the cavity mode by \SI{1}{\nano\meter}, $g^{(2)}(0)$ is the zero-delay value of the correlation function, $t_0$ is the time value corresponding to $g^{(2)}(0)$, and $\tau_C$ is the emitter lifetime.

\begin{figure}[h!]
    \centering
    \includegraphics[width=0.5\linewidth]{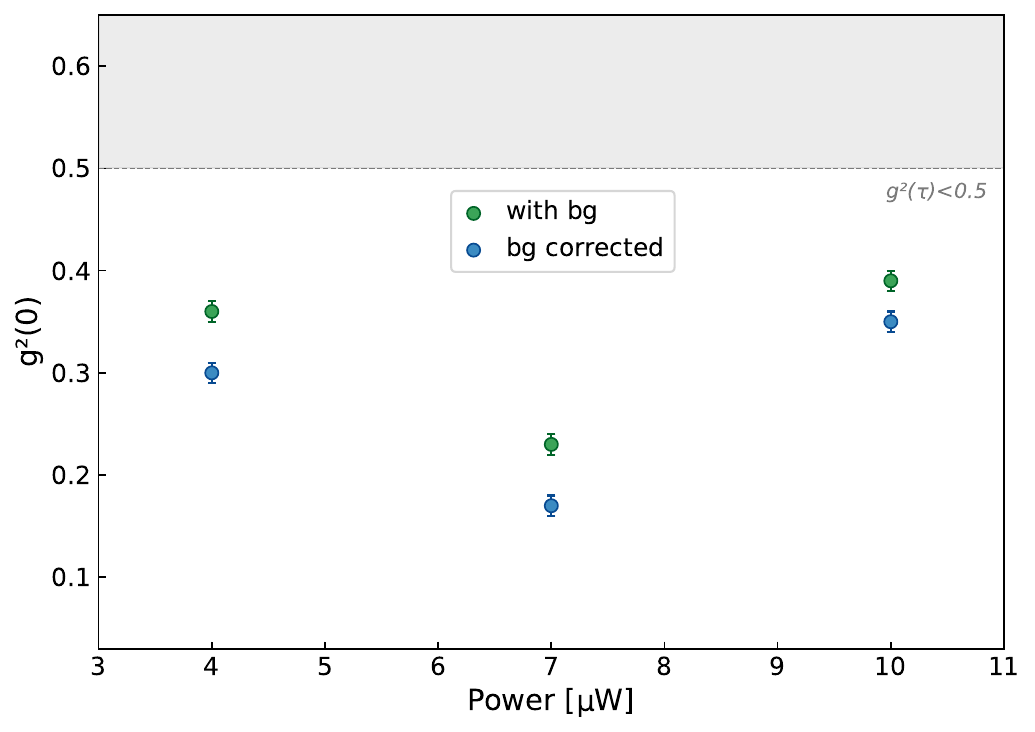}
    \caption{Non-background corrected ($I_B = 0$, green dots) and background corrected (blue dots) $g^{(2)}(0)$ values as a function of incident power.}
    \label{fig:Supp_g2(0)}
\end{figure}

The extracted $g^{(2)}(0)$ values without (green dots) -- i.e. $I_B = 0$ -- and with (blue dots) background correction are reported in \textbf{Figure \ref{fig:Supp_g2(0)}}, showing that even close to saturation power $g^{(2)}(0)$ is well below 0.5, further confirming the single-emitter nature of the investigated C-center. 

\end{document}

%% file: preamble.tex
\usepackage{amsthm}
\usepackage{mathtools}
\usepackage{physics}
\usepackage{xcolor}
\usepackage{graphicx}
\usepackage[left=23mm,right=13mm,top=35mm,columnsep=15pt]{geometry} 
\usepackage{adjustbox}
\usepackage{placeins}
\usepackage[T1]{fontenc}
\usepackage{lipsum}
\usepackage{csquotes}
\usepackage{siunitx}
\usepackage[font=small, justification=justified, labelfont=bf]{caption}

%% file: refs.bib
@article{baron2022detection,
  title={Detection of single W-centers in silicon},
  author={Baron, Yoann and Durand, Alrik and Udvarhelyi, P{\'e}ter and Herzig, Tobias and Khoury, Mario and Pezzagna, S{\'e}bastien and Meijer, Jan and Robert-Philip, Isabelle and Abbarchi, Marco and Hartmann, Jean-Michel and others},
  journal={ACS photonics},
  volume={9},
  number={7},
  pages={2337--2345},
  year={2022},
  publisher={ACS Publications}
}

@article{redjem2020single,
  title={Single artificial atoms in silicon emitting at telecom wavelengths},
  author={Redjem, Walid and Durand, Alrik and Herzig, Tobias and Benali, Abdennacer and Pezzagna, S{\'e}bastien and Meijer, Jan and Kuznetsov, A Yu and Nguyen, HS and Cueff, S{\'e}bastien and G{\'e}rard, J-M and others},
  journal={Nature Electronics},
  volume={3},
  number={12},
  pages={738--743},
  year={2020},
  publisher={Nature Publishing Group UK London}
}

@article{higginbottom2022optical,
  title={Optical observation of single spins in silicon},
  author={Higginbottom, Daniel B and Kurkjian, Alexander TK and Chartrand, Camille and Kazemi, Moein and Brunelle, Nicholas A and MacQuarrie, Evan R and Klein, James R and Lee-Hone, Nicholas R and Stacho, Jakub and Ruether, Myles and others},
  journal={Nature},
  volume={607},
  number={7918},
  pages={266--270},
  year={2022},
  publisher={Nature Publishing Group UK London}
}

@article{silkinis2025optical,
  title={Optical lineshapes of the C center in silicon from ab initio calculations: Interplay of localized modes and bulk phonons},
  author={Silkinis, Rokas and Maciaszek, Marek and {\v{Z}}alandauskas, Vytautas and Bathen, Marianne Etzelm{\"u}ller and Vines, Lasse and Alkauskas, Audrius and Razinkovas, Lukas},
  journal={Physical Review B},
  volume={111},
  number={12},
  pages={125136},
  year={2025},
  publisher={APS}
}

@article{kim2025bright,
  title={Bright purcell-enhanced single photon emission from a silicon G center},
  author={Kim, Kyu-Young and Lee, Chang-Min and Boreiri, Amirehsan and Purkayastha, Purbita and Islam, Fariba and Harper, Samuel and Kim, Je-Hyung and Waks, Edo},
  journal={Nano Letters},
  volume={25},
  number={11},
  pages={4347--4352},
  year={2025},
  publisher={ACS Publications}
}

@article{islam2023cavity,
  title={Cavity-enhanced emission from a silicon T center},
  author={Islam, Fariba and Lee, Chang-Min and Harper, Samuel and Rahaman, Mohammad Habibur and Zhao, Yuqi and Vij, Neelesh Kumar and Waks, Edo},
  journal={Nano Letters},
  volume={24},
  number={1},
  pages={319--325},
  year={2023},
  publisher={ACS Publications}
}

@article{ishikawa2009photoluminescence,
  title={Photoluminescence from triplet states of isoelectronic bound excitons at interstitial carbon-intersititial oxygen defects in silicon},
  author={Ishikawa, T and Koga, K and Itahashi, T and Vlasenko, LS and Itoh, KM},
  journal={Physica B: Condensed Matter},
  volume={404},
  number={23-24},
  pages={4552--4554},
  year={2009},
  publisher={Elsevier}
}

@article{ishikawa2011optical,
  title={Optical properties of triplet states of excitons bound to interstitial-carbon interstitial-oxygen defects in silicon},
  author={Ishikawa, T and Koga, K and Itahashi, T and Itoh, KM and Vlasenko, LS},
  journal={Phys. Rev. B},
  volume={84},
  pages={115204},
  year={2011}
}

@article{quard2025integration,
  title={Integration of color centers into silicon photonic structures},
  author={Quard, Hugo and Cueff, S{\'e}bastien and Nguyen, Hai Son and Chauvin, Nicolas and Wood, Thomas},
  journal={Applied Physics Reviews},
  volume={12},
  number={4},
  year={2025},
  publisher={AIP Publishing}
}

@article{sandholzer2026single,
  title={Single-photon emitters and spin--photon interfaces in silicon},
  author={Sandholzer, Kilian and Berkman, Ian and De{\'a}k, Peter and Errando-Herranz, Carlos and Filippatos, Petros-Panagis and Gali, Adam and Gritsch, Andreas and Reiserer, Andreas},
  journal={Applied Physics Reviews},
  volume={13},
  number={2},
  year={2026},
  publisher={AIP Publishing}
}

@article{lefaucher2023purcell,
  title={Purcell enhancement of silicon W centers in circular Bragg grating cavities},
  author={Lefaucher, Baptiste and Jager, Jean-Baptiste and Calvo, Vincent and Cache, F{\'e}lix and Durand, Alrik and Jacques, Vincent and Robert-Philip, Isabelle and Cassabois, Guillaume and Baron, Yoann and Mazen, Fr{\'e}d{\'e}ric and others},
  journal={ACS photonics},
  volume={11},
  number={1},
  pages={24--32},
  year={2023},
  publisher={ACS Publications}
}

@article{lefaucher2023cavity,
  title={Cavity-enhanced zero-phonon emission from an ensemble of G centers in a silicon-on-insulator microring},
  author={Lefaucher, Baptiste and Jager, J-B and Calvo, Vincent and Durand, Alrik and Baron, Yoann and Cache, F{\'e}lix and Jacques, Vincent and Robert-Philip, Isabelle and Cassabois, Guillaume and Herzig, Tobias and others},
  journal={Applied Physics Letters},
  volume={122},
  number={6},
  year={2023},
  publisher={AIP Publishing}
}

@article{johnston2024cavity,
  title={Cavity-coupled telecom atomic source in silicon},
  author={Johnston, Adam and Felix-Rendon, Ulises and Wong, Yu-En and Chen, Songtao},
  journal={Nature Communications},
  volume={15},
  number={1},
  pages={2350},
  year={2024},
  publisher={Nature Publishing Group UK London}
}

@article{udvarhelyi2022band,
  title={An L-band emitter with quantum memory in silicon},
  author={Udvarhelyi, P{\'e}ter and Pershin, Anton and De{\'a}k, P{\'e}ter and Gali, Adam},
  journal={npj Computational Materials},
  volume={8},
  number={1},
  pages={262},
  year={2022},
  publisher={Nature Publishing Group UK London}
}

@article{thonke1984carbon,
  title={Carbon and oxygen isotope effects in the 0.79 eV defect photoluminescence spectrum in irradiated silicon},
  author={Thonke, K and Watkins, GD and Sauer, R},
  journal={Solid state communications},
  volume={51},
  number={3},
  pages={127--130},
  year={1984},
  publisher={Elsevier}
}

@article{chartrand2018highly,
  title={Highly enriched Si 28 reveals remarkable optical linewidths and fine structure for well-known damage centers},
  author={Chartrand, C and Bergeron, L and Morse, KJ and Riemann, H and Abrosimov, NV and Becker, P and Pohl, H-J and Simmons, S and Thewalt, MLW},
  journal={Physical Review B},
  volume={98},
  number={19},
  pages={195201},
  year={2018},
  publisher={APS}
}

@article{thonke19850,
  title={0.79 eV (C line) defect in irradiated oxygen-rich silicon: excited state structure, internal strain and luminescence decay time},
  author={Thonke, K and Hangleiter, A and Wagner, J and Sauer, R},
  journal={Journal of Physics C: Solid State Physics},
  volume={18},
  number={26},
  pages={L795--L801},
  year={1985}
}

@article{wagner1984excitation,
  title={Excitation spectroscopy on the 0.79-eV (C) line defect in irradiated silicon},
  author={Wagner, J and Thonke, K and Sauer, R},
  journal={Physical Review B},
  volume={29},
  number={12},
  pages={7051},
  year={1984},
  publisher={APS}
}

@article{durand2024genuine,
  title={Genuine and faux single G centers in carbon-implanted silicon},
  author={Durand, Alrik and Baron, Yoann and Cache, F{\'e}lix and Herzig, Tobias and Khoury, Mario and Pezzagna, S{\'e}bastien and Meijer, Jan and Hartmann, Jean-Michel and Reboh, Shay and Abbarchi, Marco and others},
  journal={Physical Review B},
  volume={110},
  number={2},
  pages={L020102},
  year={2024},
  publisher={APS}
}

@article{deabreu2022waveguide,
  title={Waveguide-integrated silicon T centres},
  author={DeAbreu, Adam and Bowness, Camille and Alizadeh, Amirhossein and Chartrand, Camille and Brunelle, Nicholas A and MacQuarrie, Evan R and Lee-Hone, Nicholas R and Ruether, Myles and Kazemi, Moein and Kurkjian, ATK and others},
  journal={arXiv preprint arXiv:2209.14260},
  year={2022}
}

@article{hollenbach2022wafer,
  title={Wafer-scale nanofabrication of telecom single-photon emitters in silicon},
  author={Hollenbach, Michael and Klingner, Nico and Jagtap, Nagesh S and Bischoff, Lothar and Fowley, Ciar{\'a}n and Kentsch, Ulrich and Hlawacek, Gregor and Erbe, Artur and Abrosimov, Nikolay V and Helm, Manfred and others},
  journal={Nature Communications},
  volume={13},
  number={1},
  pages={7683},
  year={2022},
  publisher={Nature Publishing Group UK London}
}

@article{khoury2022bright,
  title={A bright future for silicon in quantum technologies},
  author={Khoury, Mario and Abbarchi, Marco},
  journal={Journal of Applied Physics},
  volume={131},
  number={20},
  year={2022},
  publisher={AIP Publishing}
}

@article{jones1973temperature,
  title={Temperature, stress, and annealing effects on the luminescence from electron-irradiated silicon},
  author={Jones, Colin E and Johnson, Eric S and Compton, W Dale and Noonan, JR and Streetman, BG},
  journal={Journal of Applied Physics},
  volume={44},
  number={12},
  pages={5402--5410},
  year={1973},
  publisher={American Institute of Physics}
}

@article{davies2006radiation,
  title={Radiation damage in silicon exposed to high-energy protons},
  author={Davies, Gordon and Hayama, Shusaku and Murin, Leonid and Krause-Rehberg, Reinhard and Bondarenko, Vladimir and Sengupta, Asmita and Davia, Cinzia and Karpenko, Anna},
  journal={Physical Review B—Condensed Matter and Materials Physics},
  volume={73},
  number={16},
  pages={165202},
  year={2006},
  publisher={APS}
}

@article{wen2025optical,
  title={Optical spin readout of a silicon color center in the telecom L-band},
  author={Wen, Shuyu and Pieplow, Gregor and Yang, Junchun and Jamshidi, Kambiz and Helm, Manfred and Luo, Jun-Wei and Schr{\"o}der, Tim and Zhou, Shengqiang and Berenc{\'e}n, Yonder},
  journal={arXiv preprint arXiv:2502.07632},
  year={2025}
}

@article{beaufils2018optical,
  title={Optical properties of an ensemble of G-centers in silicon},
  author={Beaufils, Cl{\'e}ment and Redjem, Walid and Rousseau, Emmanuel and Jacques, Vincent and Kuznetsov, A Yu and Raynaud, Christophe and Voisin, C and Benali, A and Herzig, T and Pezzagna, S and others},
  journal={Physical Review B},
  volume={97},
  number={3},
  pages={035303},
  year={2018},
  publisher={APS}
}

@article{lee2023high,
  title={High-efficiency single photon emission from a silicon T-center in a nanobeam},
  author={Lee, Chang-Min and Islam, Fariba and Harper, Samuel and Buyukkaya, Mustafa Atabey and Higginbottom, Daniel and Simmons, Stephanie and Waks, Edo},
  journal={ACS Photonics},
  volume={10},
  number={11},
  pages={3844--3849},
  year={2023},
  publisher={ACS Publications}
}

@article{andrini2024activation,
  title={Activation of telecom emitters in silicon upon ion implantation and ns pulsed laser annealing},
  author={Andrini, Greta and Zanelli, Gabriele and Ditalia Tchernij, Sviatoslav and Corte, Emilio and Nieto Hern{\'a}ndez, Elena and Verna, Alessio and Cocuzza, Matteo and Bernardi, Ettore and Virz{\`\i}, Salvatore and Traina, Paolo and others},
  journal={Communications Materials},
  volume={5},
  number={1},
  pages={47},
  year={2024},
  publisher={Nature Publishing Group UK London}
}

@article{quard2024femtosecond,
  title={Femtosecond-laser-induced creation of G and W color centers in silicon-on-insulator substrates},
  author={Quard, Hugo and Khoury, Mario and Wang, Andong and Herzig, Tobias and Meijer, Jan and Pezzagna, S{\'e}bastien and Cueff, S{\'e}bastien and Grojo, David and Abbarchi, Marco and Nguyen, Hai Son and others},
  journal={Physical Review Applied},
  volume={21},
  number={4},
  pages={044014},
  year={2024},
  publisher={APS}
}

@article{jhuria2024programmable,
  title={Programmable quantum emitter formation in silicon},
  author={Jhuria, Kaushalya and Ivanov, Vsevolod and Polley, Debanjan and Zhiyenbayev, Y and Liu, Wei and Persaud, Arun and Redjem, Walid and Qarony, Wayesh and Parajuli, Prabin and Ji, Qing and others},
  journal={Nature Communications},
  volume={15},
  number={1},
  pages={4497},
  year={2024},
  publisher={Nature Publishing Group UK London}
}

@article{zhu2026point,
  title={Point defects in semiconductors: Friends and foes for quantum technologies},
  author={Zhu, Yizhi and Zhang, Zi-Huai and Chen, Weiru and Sakib, Mashnoon Alam and Weber-Bargioni, Alexander and Griffin, Sin{\'e}ad and Raja, Archana and Sipahigil, Alp and Hautier, Geoffroy},
  journal={MRS Bulletin},
  pages={1--15},
  year={2026},
  publisher={Springer}
}

@article{bohnert1993transient,
  title={Transient characteristics of isoelectronic bound excitons at hole-attractive defects in silicon: The C (0.79 eV), P (0.767 eV), and H (0.926 eV) lines},
  author={Bohnert, G and Weronek, K and Hangleiter, A},
  journal={Physical Review B},
  volume={48},
  number={20},
  pages={14973},
  year={1993},
  publisher={APS}
}

@article{islam2026spin,
  title={Spin-photon qubits for scalable quantum network},
  author={Islam, Md Sakibul and Singh, Kuldeep and Zhao, Yunhe and Singh, Nitesh and Qarony, Wayesh},
  journal={Light: Science \& Applications},
  volume={15},
  number={1},
  pages={301},
  year={2026},
  publisher={Nature Publishing Group UK London}
}

@article{aharonovich2010photophysics,
  title={Photophysics of chromium-related diamond single-photon emitters},
  author={Aharonovich, I and Castelletto, Stefania and Simpson, DA and Greentree, AD and Prawer, Steven},
  journal={Physical Review A—Atomic, Molecular, and Optical Physics},
  volume={81},
  number={4},
  pages={043813},
  year={2010},
  publisher={APS}
}

@article{kim2013ultrafast,
  title={Ultrafast single photon emitting quantum photonic structures based on a nano-obelisk},
  author={Kim, Je-Hyung and Ko, Young-Ho and Gong, Su-Hyun and Ko, Suk-Min and Cho, Yong-Hoon},
  journal={Scientific reports},
  volume={3},
  number={1},
  pages={2150},
  year={2013},
  publisher={Nature Publishing Group UK London}
}

@inproceedings{biswas2025single,
  title={Single Photon Emission from InAs/GaAs Quantum Dot Embedded in High-Efficiency Tapered Nanobeam Cavity},
  author={Biswas, Abhijit and Bracker, Allan S and Waks, Edo},
  booktitle={2025 Conference on Lasers and Electro-Optics (CLEO)},
  pages={1--2},
  year={2025},
  organization={IEEE}
}

@article{gritsch2023purcell,
  title={Purcell enhancement of single-photon emitters in silicon},
  author={Gritsch, Andreas and Ulanowski, Alexander and Reiserer, Andreas},
  journal={Optica},
  volume={10},
  number={6},
  pages={783--789},
  year={2023},
  publisher={Optica Publishing Group}
}
